# Quantum theory of vortex lattice state in a rotating Bose-Einstein condensate

Aranya B Bhattacherjee

Department of Physics, A.R.S.D College, University of Delhi (South Campus), Dhaula Kuan, New Delhi-110 021, India.

**Abstract:** We study system of large number of singly quantized vortices in a rotating Bose-Einstein condensate. Analogous to the Meissner effect in superconductors, we show that the vector potential due to the external rotational field can be tuned to cancel the vector potential due to the Magnus field, resulting in a zero average angular momentum and a shear modulus of the vortex lattice. The vortex lattice state exhibits two states, namely, an elastic state and a plastic state. A clear distinction between these states is controlled by the filling fraction $v = N/N_V$, which is the ratio of the number of bosons ($N$) to the average number of vortices ($N_V$).

# Introduction

Quantized vortices play an important role in the behaviour of superfluids [1,2]. The vortices provide a mechanism for the decay of superfluid currents in a ring. Vortices in a superfluid formed due to rapid rotation have an associated rigidity. In the recent years, it has become possible to achieve such a vortex lattice state in a rapidly rotating Bose Einstein condensate (BEC)[3-6].

There are strong analogies between the behaviour of electrons in strong magnetic fields and of vortices in superfluids. These analogies led to the prediction that quantum-Hall like properties should emerge in rapidly rotating atomic BEC's [7,8]. In particular, the single particle energy states organize into Landau levels, and if interactions are weaker than the cyclotron energy, primarily the near degenerate states of the lowest Landau level (LLL) are occupied. In this regime a drastic decrease of the lattice's elastic shear strength takes place. The elastic shear modulus predicted by Baym [9,10] decreases with increasing rotation rate from its value in the "stiff" Thomas-Fermi (TF) limit ($C_2^{TF} = n\hbar\Omega/8$, where $n$ is the BEC number density and $\Omega$ is the rotation rate of the external trap) to its value in the mean-field quantum Hall regime ($C_2^{LLL} = 0.16\Gamma_{LLL} C_2^{TF}, \Gamma_{LLL} = \mu/2\hbar\Omega$). In this work, we study the dynamics of a large number of singly quantized vortices in a rotating BEC. In particular, we study the effective Hamiltonian for the vortex degree of freedom, motivated by an analogy between the vector potential associated with the Magnus force acting on a vortex moving in a two-dimensional neutral superfluid and the Chern-Simons vector potential associated with the internal degrees of freedom of a quantum Hall liquid. The vector potential associated with the coriolis force is equivalent to the vector potential of the Lorentz force acting on a charged particle in a magnetic field. This picture is different from the previous analogies between vortices in superfluids and charged particles in magnetic fields. Based on this picture, we show an important effect, analogous to the Meissner effect in superconductors, due to competition between external and internal degrees of freedom.

# Effective theory of quantum vortex state

We consider a large number of vortices in a rotating gas of bosons confined in a trap at temperatures well below the Kosterlitz-Thouless transition temperature given by $k_B T_C^{KT} = \hbar^2 \rho_s / 4m^2$, where $\rho_s$ is the mass density and $m$ is the mass of the single bosonic atom. The BEC is described by a repulsive short-range interaction $V(\vec{r}) = g_{2D} \delta^2(\vec{r})$, with the 2D interaction parameter $g_{2D} = \sqrt{8\pi} \hbar^2 a_s / m a_z$ [11], where $a_s$ is the s-wave scattering length and $a_z = \sqrt{\hbar / m\omega_z}$ is the axial oscillator length. $\omega_z$ is the axial trapping frequency. One should note that this value of the 2D interaction parameter becomes incorrect at extremely low densities and becomes independent of the scattering length $a_s$ [11]. The exact value of the 2D interaction parameter upto logarithmic accuracy is $g_{2D} = \dfrac{\sqrt{8\pi} \hbar^2}{m} \dfrac{1}{a_z/a_s + (1/\sqrt{2\pi}) \ln(1/\pi q^2 a_z^2)}$. The 2D interaction parameter depends on $q = \left(2m|\mu|/\hbar^2\right)^{1/2}$ ($\mu$ is the chemical potential) and, hence on the condensate density. In the limit $a_z \gg |a_s|$ the logarithmic term is not important, and the 2D interaction parameter becomes density independent. We will always be working in the limit $a_z \gg |a_s|$. The superfluid forms a triangular lattice of quantized vortices (carrying the angular momentum of the system) rotating as a solid body at angular velocity $\Omega$. We assume that at low temperatures, the scattering of thermal excitations by vortices does not affect the vortex dynamics. The situation considered then corresponds to considering a vortex as a point particle moving under the influence of the Magnus force.

The vortex motion is governed by a Hamiltonian corresponding to one of point particles, with a charge equal to the quantum of circulation $\kappa = h/m$, interacting with electromagnetic fields. For a large collection of vortices $N_V \gg 1$ in a frame rotating with angular velocity $\Omega$, the vortex Hamiltonian reads:

$$H_V = \sum_{i=1}^{N_V} \frac{(\vec{P}_i - \kappa \vec{a}_i)^2}{2m_v} - \Omega(\vec{X}_i \times \vec{P}_i)_z + V_{ij}(\vec{X}_i - \vec{X}_j) \tag{1}$$

The first term in Eqn.(1) is the kinetic energy of the vortices, the second term is a result of centrifugal and coriolis forces on vortices and the last term $V_{ij}(\vec{X}_i - \vec{X}_j) = -\frac{\rho_S \kappa^2}{2\pi} \sum_{i<j}^{N_V} \ln\left|\frac{\vec{X}_i - \vec{X}_j}{\xi}\right|$ is the repulsive logarithmic Coulomb interaction of point vortices. $\xi$ is the coherence length and is $\xi = \sqrt{\hbar^2 / 2g_{2D}\rho_S}$. The effective vortex mass is $m_v = \pi \rho_S \xi^2$ [11]. The pseudo vector potential due to the Magnus force is $a_i^a = \frac{1}{2}\rho_S \varepsilon^{ab} X_b^i$ $(a,b = x, y)$ [12]. At very high rotations the BEC approaches a quasi-2D regime due to centrifugal force [13]. We can rewrite eqn.(1) as

$$H_V = \sum_{i=1}^{N_V} \frac{(\vec{P}_i - \kappa \vec{A}_i - \kappa \vec{a}_i)^2}{2m_v} + \kappa(A_0 + a_0) + V_{ij}(\vec{X}_i - \vec{X}_j) \tag{2}$$

The vector potentials are,

$$\vec{A}_i = \frac{m_v \Omega}{\kappa}(-\hat{i}y_i + \hat{j}x_i),$$

$$\vec{a}_i = \frac{\rho_S}{2}(\hat{i}y_i - \hat{j}x_i), \tag{2a}$$

The scalar potential are

$$a_0 = \frac{\Omega \rho_S}{2}(x_i^2 + y_i^2),$$

$$A_0 = -\frac{\Omega^2 m_v}{2\kappa}(x_i^2 + y_i^2), \tag{2b}$$

We define the average value of the vortex flux $<\vec{J}_{vor}>$ as

$$<\vec{J}_{vor}> = \sum_i \vec{\nabla}_i \times \kappa\{<\vec{A}_i> + <\vec{a}_i>\} = N_V\{2\Omega m_v - \kappa <\rho_s>\}\hat{z} \qquad (3)$$

Where $<\rho_s>$ is the average value of $\rho_s$. $<\vec{J}_{vor}>$ vanishes when $\{2\Omega m_v - \kappa <\rho_s>\} = 0$. This condition is expressed as

$$v = \sqrt{\frac{\pi}{32}\frac{a_z}{a_s}} \qquad (3a)$$

The parameters of the JILA experiments [13] give the filling fraction $v = 293$. Later in this paper, we show that the elastic shear modulus $C_2$ is proportional to $<\vec{J}_{vor}>$. Analogous to the Meissner effect in superconductors, the external rotation $\Omega$ together with appropriate trap parameters can be tuned to make the average angular momentum of the system $<\vec{L}_z> = \sum_i <\vec{J}_{vor}>(x_i^2 + y_i^2)$ zero. Experimentally this effect can be observed by measuring the lowest order azimuthally symmetric Tkachenko lattice mode [13]. Its frequency $\omega_{1,0} \propto \sqrt{C_2}$ is expected to go to zero at $v = \sqrt{\frac{\pi}{32}\frac{a_z}{a_s}}$. At higher rotations, ($v < \sqrt{\frac{\pi}{32}\frac{a_z}{a_s}}$), interestingly, the average angular momentum $<\vec{L}_z>$ reverses its direction.

In second quantization, the Hamiltonian of Eqn.(2) reads

$$H = \int d^2\vec{X}\phi^+(\vec{X})\left\{\frac{1}{2m_v}\left[-i\hbar\vec{\nabla} + \kappa\vec{A}(\vec{X}) + \kappa\vec{a}(\vec{X})\right]^2 + \kappa\left[A_0(\vec{X}) + a_0(\vec{X}) - \tilde{\mu}\right]\right\}\phi(\vec{X})$$
$$+ \frac{1}{2}\int d^2\vec{X}\int d^2\vec{X}'\rho(\vec{X})V(\vec{X} - \vec{X}')\rho(\vec{X}')$$

$$(4)$$

Here $\phi^+(\bar{X})$ and $\phi(\bar{X})$ are the Bosonic creation and annihilation operators, $[\phi(\bar{X}), \phi^+(\bar{X}')] = \delta(\bar{X} - \bar{X}')$ and $\rho(\bar{X}) = \phi^+(\bar{X})\phi(\bar{X})$ is the particle density. $\tilde{\mu}$ is the chemical potential. Under the presence of the external rotational field $A_\mu$, the path integral of the system reads

$$Z[A_\mu] = i \int Da_\mu D\phi \exp[iS_\phi(A_\mu + a_\mu, \phi)], \mu = x, y, 0 \tag{5}$$

Where $S_\phi$ is given by

$$S_\phi[A_\mu + a_\mu, \phi] = \int dt \int d^2\bar{X} \left\{ \begin{array}{l} \phi^+(i\hbar\partial_t - \kappa[A_0(\bar{X}) + a_0(\bar{X}) - \tilde{\mu}])\phi \\ -\dfrac{1}{2m_v} |(-i\hbar\bar{\nabla} + \kappa(\bar{A} + \bar{a}))\phi|^2 \end{array} \right\} \tag{6}$$

$$-\frac{1}{2}\int dt \int d^2\bar{X} \int d^2\bar{X}' \rho(\bar{X}) V(\bar{X} - \bar{X}') \rho(\bar{X}')$$

It is to be noted that the above action is valid under the Coulomb gauge condition i.e. $\bar{\nabla}_i \cdot (\bar{A}_i + \bar{a}_i) = 0$ and the system is not gauge invariant. By representing the bosonic field $\phi(\bar{X})$ by its phase $\theta(\bar{X})$ and amplitude $\rho(\bar{X})$

$$\phi(\bar{X}) = \sqrt{\rho} \exp[i\theta(\bar{X})] \tag{7}$$

$$S_\theta[A_\mu + a_\mu, \theta, \rho] = \int dt \int d^2\bar{X} \left\{ \begin{array}{l} \rho(-\hbar\partial_t\theta - \kappa[A_0(\bar{X}) + a_0(\bar{X}) - \tilde{\mu}]) \\ -\dfrac{\rho}{2m_v}|(\hbar\bar{\nabla}\theta + \kappa(\bar{A} + \bar{a}))|^2 - \dfrac{\rho^{-1}}{8m_v}(\hbar\bar{\nabla}\rho)^2 \end{array} \right\} \tag{8}$$

$$-\frac{1}{2}\int dt \int d^2\bar{X} \int d^2\bar{X}' \rho(\bar{X}) V(\bar{X} - \bar{X}') \rho(\bar{X}')$$

Assuming that the saddle point solution is independent of space and time, we obtain the condition

$$\rho_0 = \frac{\tilde{\mu}}{\tilde{V}(0)} > 0 \tag{9}$$

$\tilde{V}(k)$ is the Fourier transform of $V(\vec{X})$. We now expand $\rho$ as a function of $\rho_0$ and $\delta\rho$ and expand the action up to second order in $\delta\rho$ and $\partial_\mu\theta + \kappa(A_\mu + a_\mu)$.

$$S_\theta[A_\mu + a_\mu, \theta] = \int dt \int d^2\vec{X} \begin{Bmatrix} \delta\rho(-\hbar\partial_t\theta - \kappa[A_0(\vec{X}) + a_0(\vec{X})]) \\ -\frac{\rho_0}{2m_v}\left|(\hbar\vec{\nabla}\theta + \kappa(\vec{A} + \vec{a}))\right|^2 - \frac{\rho_0^{-1}}{8m_v}(\hbar\vec{\nabla}\delta\rho)^2 \end{Bmatrix}$$
$$-\frac{1}{2}\int dt \int d^2\vec{X} \int d^2\vec{X}' \delta\rho(\vec{X}) V(\vec{X} - \vec{X}') \delta\rho(\vec{X}') \tag{10}$$

In the small $\vec{k}$ limit the term proportional to $(\vec{\nabla}\delta\rho)^2$ makes a small contribution compared with the interaction of the third term, we ignore it. After Fourier transformation and $\delta\rho$ integration, we obtain

$$S_\theta[A_\mu + a_\mu, \theta] = \sum_\omega \sum_{\vec{k}} \begin{Bmatrix} \frac{1}{2\tilde{V}(\vec{k})}[-i\hbar\omega\theta(-k) - \kappa(A_0(-k) + a_0(-k))] \\ \times[i\hbar\omega\theta(k) - \kappa(A_0(k) + a_0(k))] \\ -\frac{\rho_0}{2m_v}[-i\hbar\vec{k}\theta(-k) + \kappa(\vec{A}(-k) + \vec{a}(-k))] \\ \times[i\hbar\vec{k}\theta(k) + \kappa(\vec{A}(k) + \vec{a}(k))] \end{Bmatrix} \tag{11}$$

Writing $\kappa(\vec{A}(k) + \vec{a}(k)) = \delta\vec{a}(k)$ and $\kappa(A_0(k) + a_0(k)) = \delta a_0(k)$ and doing the $\theta$ integration, we obtain

$$S_{eff} = \sum_{k,\mu,\nu} \frac{1}{2}\pi_{\mu\nu}(k)\delta a_\mu(-k)\delta a_\nu(k) \tag{12}$$

$$\pi_{00}(k) = \frac{1}{\tilde{V}(\vec{k})} \left\{ \frac{-\left(\frac{\rho_0}{2m_v}\right)|\hbar\vec{k}|^2}{\left[\frac{\hbar^2\omega^2}{2\vec{V}(\vec{k})} - \left(\frac{\rho_0}{2m_v}\right)|\hbar\vec{k}|^2\right]} \right\} \qquad (13)$$

$$\pi_{\alpha\beta}(k) = -\frac{\rho_0}{m_v} \left\{ \delta_{\alpha\beta} + \frac{\left(\frac{\rho_0}{2m_v}\right)\hbar^2 k_\alpha k_\beta}{\left[\frac{\hbar^2\omega^2}{2\vec{V}(\vec{k})} - \left(\frac{\rho_0}{2m_v}\right)|\hbar\vec{k}|^2\right]} \right\} \qquad (14)$$

$$\pi_{0\alpha}(k) = \pi_{\alpha 0}(k) = -\frac{\rho_0 \hbar^2 \omega k_{\alpha 0}}{2m_v \tilde{V}(\vec{k})} \left\{ \frac{1}{\left[\frac{\hbar^2\omega^2}{2\vec{V}(\vec{k})} - \left(\frac{\rho_0}{2m_v}\right)|\hbar\vec{k}|^2\right]} \right\} \qquad (15)$$

Generally speaking, $S_{eff}$ represents the linear response of the vortex lattice to the external rotational field $A_\mu$ and internal degree of freedom (due to the magnus force) $a_\mu$ that is correlated to the fluctuations of the particle mass density $\rho_s(\vec{X})$. We represent it as the sum of the average value $<a_\mu>$ and fluctuations $\delta a_\mu$ around it, $<a_\mu> + \delta a_\mu$. We write for $A_\mu$, the sum of $<A_\mu>$ corresponding to a constant rotation $\Omega$ along the $z$ direction and an infinitesimal test field $\delta A_\mu$ that is introduced to study the linear response of the system, $<A_\mu> + \delta A_\mu$. Analogous to the Meissner effect in superconductors, the external degree of freedom $A_\mu$ can be tuned to cancel the internal degree of freedom $a_\mu$. $\pi_{00}(k)$ is the longitudinal response of the system. $\pi_{\alpha\beta}(k)$ is the transverse response. Interestingly as $\vec{k} \to 0$, $\pi_{00}(k) \to 0$ but $\pi_{\alpha\beta}(k) \to -\rho_0/m_v$. The transverse response is better understood in terms of energy increase of the vortex lattice due to change in $(\vec{A}+\vec{a}) = \delta\vec{a}$. In terms of the lattice displacement vector $\vec{\varepsilon}(\vec{X}) = (-\hat{i}y + \hat{j}x)$,

$$\delta\vec{a}(\vec{X}) = \left[m_v\Omega - \frac{\kappa\rho_s}{2}\right]\vec{\varepsilon}(\vec{X}) \tag{16}$$

We ask what is the work done by an effective force field $\delta F_{eff} = \left[m_v\Omega^2 - \frac{\kappa\rho_s\Omega}{2}\right](\partial_x\varepsilon_y - \partial_y\varepsilon_x) = \left[m_v\Omega^2 - \frac{\kappa\rho_s\Omega}{2}\right]\delta\varepsilon$? The effective force field is the sum of the magnus force and the pseudo force. Written in the action, the energy increase is

$$E(\vec{X}) = -\frac{1}{2}\left[m_v\Omega^2 - \frac{\kappa\rho_s\Omega}{2}\right]\int d\vec{X}(\delta\varepsilon)^2 \tag{17}$$

Thus for small rotations ($v > \sqrt{\frac{\pi}{32}\frac{a_z}{a_s}}$) as the energy increases, the vortex lattice tries to suppress the rotation $\Omega$. The magnus force acts as an elastic restoring force for small rotations. From eqn.(12) and together with the definition of shear modulus of ref.[10], we identify the shear modulus as

$$C_2 = \alpha\left\{n\hbar\Omega - \frac{n_v\hbar\Omega a_z}{a_s}\sqrt{\frac{\pi}{32}}\right\} \tag{18}$$

Here $\alpha$ is a constant, which depends on the regime. In the Thomas-Fermi regime, $\alpha = 1/8$. At $v = \sqrt{\frac{\pi}{32}\frac{a_z}{a_s}}$, the shear modulus vanishes. This is to be compared with the result of Baym [10], $C_2 \to 0$ as $\Omega \to \omega_\perp$ (radial trap frequency). At higher rotations $v < \sqrt{\frac{\pi}{32}\frac{a_z}{a_s}}$, $C_2$ becomes negative and the vortex lattice yields to the rotation $\Omega$ and enters a plastic state. As the cold cloud deviates from its 2D shape (i.e. $a_z > a_\perp$), $C_2 \to 0$ for $\Omega < \omega_\perp$.

## Conclusion

In conclusion, we have shown that for a system comprising of a large number of vortices in a rotating Bose-Einstein condensate, there is a competition between the internal degree of freedom ($\bar{a}$) and the external degree of freedom ($\bar{A}$). The vector potential $\bar{A}$ can be tuned to cancel $\bar{a}$ resulting in a zero average angular momentum and a zero shear modulus. This effect is analogous to the Meissner effect in superconductors and occurs at a critical value of the filling fraction $\nu = \nu_c$, $\nu_c = \sqrt{\dfrac{\pi}{32} \dfrac{a_z}{a_s}}$. The vortex lattice state exhibits both an elastic and plastic state. A clear distinction between these states is controlled by the boson particle density $n$ and vortex density $n_V$. Vortex elastic state appears for $\nu \geq \nu_c$ and vortex plastic state appears for $\nu \leq \nu_c$. A transition from one state to the other is characterized by a change in the direction of the average angular momentum.